\documentclass{article}

\usepackage{PRIMEarxiv}

\usepackage[utf8]{inputenc} 
\usepackage[T1]{fontenc}    
\usepackage{hyperref}       
\usepackage{url}            
\usepackage{booktabs}       
\usepackage{amsfonts}       
\usepackage{nicefrac}       
\usepackage{microtype}      
\usepackage{lipsum}
\usepackage{fancyhdr}       
\usepackage{graphicx}       
\graphicspath{{media/}}     

\title{Identifying Damage-Sensitive Spatial Vibration Characteristics of Bridges from Widespread Smartphone Data
}

\author{
  Liam Cronin \\
  Lehigh University\\
  \texttt{lmc219@lehigh.edu} \\
  \And
  Soheil Sadeghi Eshkevari \\
  MIT Senseable City Lab \\
  \texttt{ssadeghi@mit.edu} \\
  \And
  Thomas J. Matarazzo \\
  USMA West Point\\
  \texttt{thomas.matarazzo@westpoint.edu} \\
  \And
  Sebastiano Milardo \\
  MIT Senseable City Lab \\
  \texttt{milardo@mit.edu} \\
  \And
  Iman Dabbaghchian \\
  Lehigh University \\
  \texttt{imd220@lehigh.edu} \\
  \And
  Paolo Santi \\
  MIT Senseable City Lab \\
  \texttt{psanti@mit.edu} \\
  \And
  Shamim N. Pakzad \\
  Lehigh University \\
  \texttt{pakzad@lehigh.edu} \\
  \And
  Carlo Ratti \\
  MIT Senseable City Lab \\
  \texttt{ratti@mit.edu} \\
  }

\begin{document}
\maketitle

\begin{abstract}
The knowledge gap in the expected and actual conditions of bridges has created worldwide deficits in infrastructure service and funding challenges. Despite rapid advances over the past four decades, sensing technology is still not a part of bridge inspection protocols. Every time a vehicle with a mobile device passes over a bridge, there is an opportunity to capture potentially important structural response information at a very low cost. Prior work has shown how bridge modal frequencies can be accurately determined with crowdsourced smartphone-vehicle trip (SVT) data in real-world settings. However, modal frequencies provide very limited insight on the structural health conditions of the bridge. Here, we present a novel method to extract spatial vibration characteristics of real bridges, namely, absolute mode shapes, from crowdsourced SVT data. These characteristics have a demonstrable sensitivity to structural damage and provide superior, yet complementary, indicators of bridge condition. Furthermore, they are useful in the development of accurate mathematical models of the structural system and help reconcile the differences between models and real systems. We demonstrate successful applications on four very different bridges, with span lengths ranging from about 30 to 1300 meters, collectively representing about one quarter of bridges in the US. Supplementary work applies this computational approach to accurately detect simulated bridge damage entirely from crowdsourced SVT data in an unprecedentedly timely fashion. The results presented in this article open the way towards large-scale crowdsourced monitoring of bridge infrastructure.
\end{abstract}


\section*{Introduction}

Ubiquitous smartphone devices have normalized the distributed collection of scientific data in everyday life. Sensor arrays in the modern smartphones enable unprecedented measurements on an individual's activities. In contrast to increasing digital capabilities, modern society faces significant infrastructure deficits. Knowledge gaps regarding the conditions of infrastructure systems have created vulnerabilities to sudden and unexpected losses in service, and have ultimately produced an infrastructure-funding gap. For instance, at current investment levels, it would take about 50 years for the U.S. departments of transportation (DOTs) to resolve outstanding repairs\footnote{This projection is conservative as it does not include any new bridge maintenance issues that would arise.}. The total 10-year infrastructure-funding gap for roadways and transit is over one trillion USD \cite{american2016failure}. Furthermore, the exposure of infrastructure to natural disasters amplifies uncertainties in asset management. In regions with high risks of multiple natural hazards, e.g., earthquakes, coastal flooding, cyclones, etc., transportation infrastructure are subject to significant losses. For example, road bridges in China account for about 29\% of the expected annual damage caused by natural hazards \cite{koks2019global}. Bridges and other road infrastructure are vulnerable to the rising rates of natural disasters and extreme events in response to climate change and rapid growth in the global human footprint \cite{van2006impacts,mitchell2006extreme,banholzer2014impact}. In particular, flooding and other hydraulic events are a leading cause of bridge failures in the U.S. \cite{cook2015bridge}, the U.K. \cite{van2014flood}, India \cite{garg2022analysis}, and other countries.
This emphasizes the importance of incorporating accurate models of bridges and their exposure to extreme events in life-cycle analyses \cite{jeong2018bridge}.

Modern structural health monitoring (SHM) techniques are highly capable of determining important physical characteristics of bridges based on sensor data. SHM encompasses a wide range of services such as modal identification, damage identification and localization, digital twin modelling, risk quantification, and disaster response \cite{lynch2007overview,farrar2012structural}. Advances in sensing and actuation have led to applications utilizing computer-vision and robotics, e.g., automatic crack detection, drone-based inspections, etc., which are designed to improve the retrieval of structural condition information \cite{sanayei2015automated,khaloo2018unmanned,momtaz2022color}. However, the costs associated with implementing these techniques (even in their simplest forms) have proven to be unattainable for the vast majority of bridge owners. Sensing technology is not a part of routine bridge inspections: U.S. National Bridge Inspection Standards only require that each bridge is inspected \emph{visually} in 24-month intervals. A widespread need for monitoring approaches that are accurate, easy-to-implement, and cost-effective has helped spark interest in low-cost alternatives, such as the use of mobile sensor networks \cite{lin2005use} and smartphones \cite{feng2015citizen} in SHM. 

In the last two decades, researchers  \cite{yang2004extracting,lin2005use,yang2009extracting,siringoringo2012estimating,zhang2012damage,feng2015citizen,mcgetrick2017implementation,yang2020vehicle, sitton2020bridge, sitton2020frequency} have established the advantages that mobile sensor networks have over traditional ``stationary'' ones in measuring bridge vibrations. Two key findings for mobile sensor networks are (i) few devices are needed to determine structural dynamical properties, which enables widespread bridge monitoring at a lower cost; and (ii) mobile sensors efficiently capture dense spatial vibration information, e.g., high-resolution structural mode shapes \cite{marulanda2016modal,matarazzo2016structural,matarazzo2016truncated,matarazzo2018scalable}. Recent work presented multiple real-world applications in which crowdsourced smartphone-vehicle-trip (SVT) data were used to accurately determine modal properties of two bridges \cite{matarazzo2020crowdsourcing}. These results emphasized the unique and significant advantages of crowdsourced data. ``Pre-existing'' mobile sensor networks \cite{OKeeffe12752} such as ridesharing data, municipal transit data, etc., provide an unprecedented potential for high velocity and large-scale data streams. Modal frequency identification is an essential first step in monitoring the dynamics and condition of a bridge. While modal frequencies have been successfully used to identify structural damage in real-world applications, certain modal frequencies, e.g., lower modes, can be insensitive to damage and simultaneously sensitive to normal environmental changes \cite{askegaard1988long,peeters2001one,peeters2001vibration, liang2018frequency, ralbovsky2010frequency, kim2007vibration, jin2016damage, fan2011vibration}, which can reduce their effectiveness as a damage-sensitive feature. A robust and accurate monitoring system must incorporate several layers of information into a condition report such as  environmental data, dynamical properties, and damage-sensitive features.

\begin{figure*}
\centering
\includegraphics[width=0.74\paperwidth]{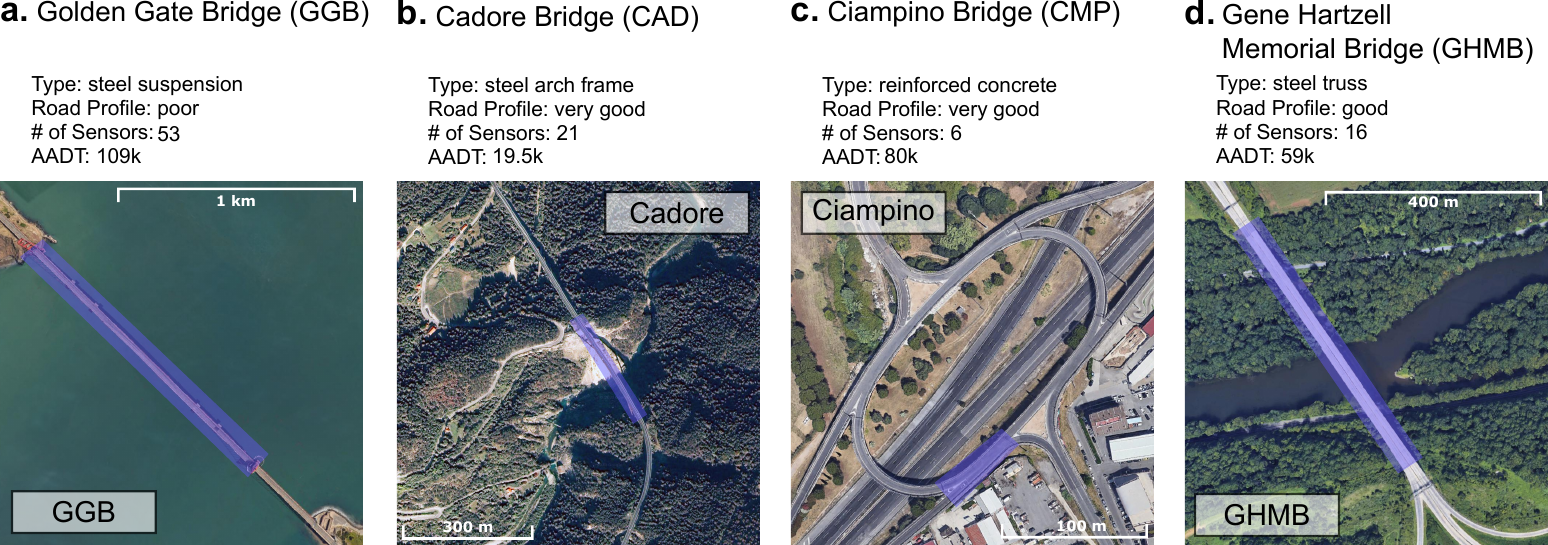}
\caption{Aerial views and general information for the four case studies. The monitored sections are highlighted in the photos. These bridges are monitored with traditional fixed sensor networks, and results are used as a baseline to compare with the mobile sensing campaign. In summary, these bridges display a wide range of characteristics with span-lengths varying from 56m to 1,280m and consist of four distinct structural systems.}
\label{fig:bridges}
\end{figure*}

Spatial vibration information, such as mode shapes, are both effective damage-sensitive features \emph{and} can be captured efficiently by mobile sensors. Structural mode shapes and their curvatures are sensitive features to local and global damage while less sensitive to environmental variations \cite{farrar1997system}. Over time, as structural damage develops, the deviations in mode-shape-based metrics \cite{shi2000damage,lee2005neural,hu2006statistical} can lead to the identification of both the presence and location of the damage. Similarly, signal processing techniques, e.g., wavelet transforms, have been used to identify mode shape discontinuities and attributed them to local damage \cite{liew1998application,hong2002damage,douka2004crack,chang2005detection,poudel2007wavelet,tan2017wavelet}. For these reasons, mode shapes and mode shape curvatures are widely studied and used for damage detection and localization \cite{pandey1991damage,wahab1999damage,kim2006damage,feng2016output,shokrani2018use}.

This paper proposes a method for identifying absolute value mode shapes (AMS) of bridge structures exclusively from crowdsourced SVT data. The method is validated throughout five real-world applications on four distinct bridges, Figure \ref{fig:bridges}: The Golden Gate Bridge (USA), the Cadore Bridge (Italy), the Ciampino Bridge (Italy), and the Gene Hartzell Memorial Bridge (USA). These bridges have distinctly different locations, designs, and traffic volumes. Notably, the lengths of the largest span vary from 30 m to 1.3 km; a range that represents 32\% of US bridges.

The proposed method has three key benefits: (1) it is entirely based on SVT data which can be sourced from common smartphones and does not require supplementary sensory or GIS information, (2) with ever-expanding smartphone penetration rates, it can streamline up-to-date AMS estimations on a regular basis, which is critical for long-term, reference-based  health monitoring, and (3) the spatial resolution of the AMS is high, which enables immediate applications to broader SHM services such as damage identification and localization. These features can be configured with a software-as-service (SaS) system for automated data collection, preprocessing, and cloud-based storage, that can provide tools for near-real-time condition assessments.

\section*{SVT Aggregation and AMS Identification Method}

\begin{figure}[ht!]
\centering
\includegraphics{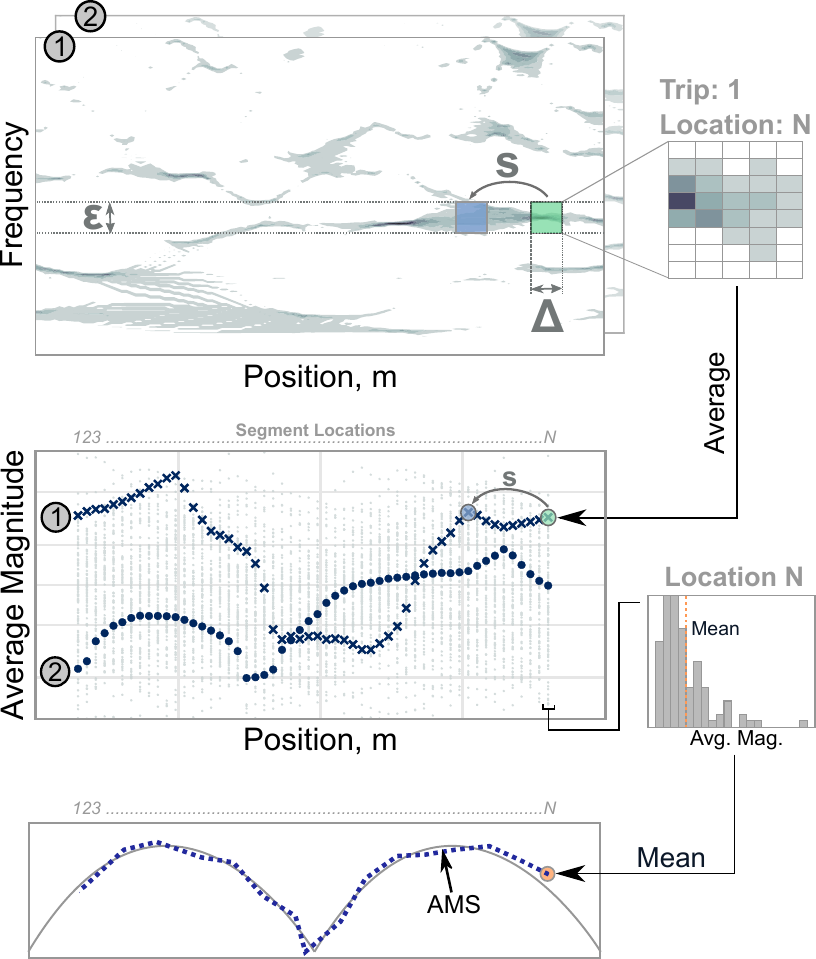}
\caption{Overview of the methodology: the synchrosqueezed wavelet transform converts each acceleration time series to the time-frequency domain, and the instantaneous magnitudes are calculated. The bridge is divided into overlapping segments; the width ($\Delta$) and spatial stride ($S$) are two parameters of the method. Averaging the magnitudes in each segment for each trip at the modal frequencies yields a distribution of magnitudes at each location, the mean of which is the AMS estimate. Including a small bandwidth ($\epsilon$) around the modal frequency in the averaging process leads to robust results on noisy datasets.  Detailed information is given in Materials and Methods.}
\label{fig:method}
\end{figure}

The process for identifying AMS is represented graphically in Figure \ref{fig:method}. A detailed explanation of the proposed methodology for AMS identification is given in Materials and Methods section Table \ref{tbl:alg}. After basic preprocessing steps, the synchrosqueezed wavelet transform is applied to the SVT acceleration signals and the values are mapped to frequency-space grids (see top panel in Figure \ref{fig:method}), which are defined by the width ($\Delta$) and stride ($S$) parameters (scalar values).

This process is repeated for all signals in a batch of SVT data for which the individual frequency-space grids are aggregated. Narrow frequency bands, which are defined based on prior knowledge of modal frequencies\footnote{It is key to emphasize that this method requires estimates of natural frequencies. An approach for determining modal frequencies from SVT data is reviewed in the SI along with validations  on the presented case studies \cite{matarazzo2020crowdsourcing}.}, are applied to the aggregated frequency-space grid to plot the absolute signal power at a central frequency as a function of location (see middle panel in Figure \ref{fig:method}) for every trip. The mean value over all trips in the batch is extracted as the AMS for the modal frequency (see bottom panel in Figure \ref{fig:method}). At this stage, the spatial resolution of the AMS can be adjusted by the user, but it is generally bounded by the sampling rate and the vehicle speed.

 The accuracy of the AMS is evaluated using modal assurance criteria (MAC), which produces a value between $0$ and $1$\footnote{Estimated mode shapes with MAC values above $0.90$ indicate a high consistency with the reference shape and are regarded as accurate} that measures the quality of an estimated mode shape by comparing it with the reference mode shape \cite{allemang1982correlation}.

\section*{Results}

\begin{figure*}[ht]
\centering
\includegraphics[width=155mm]{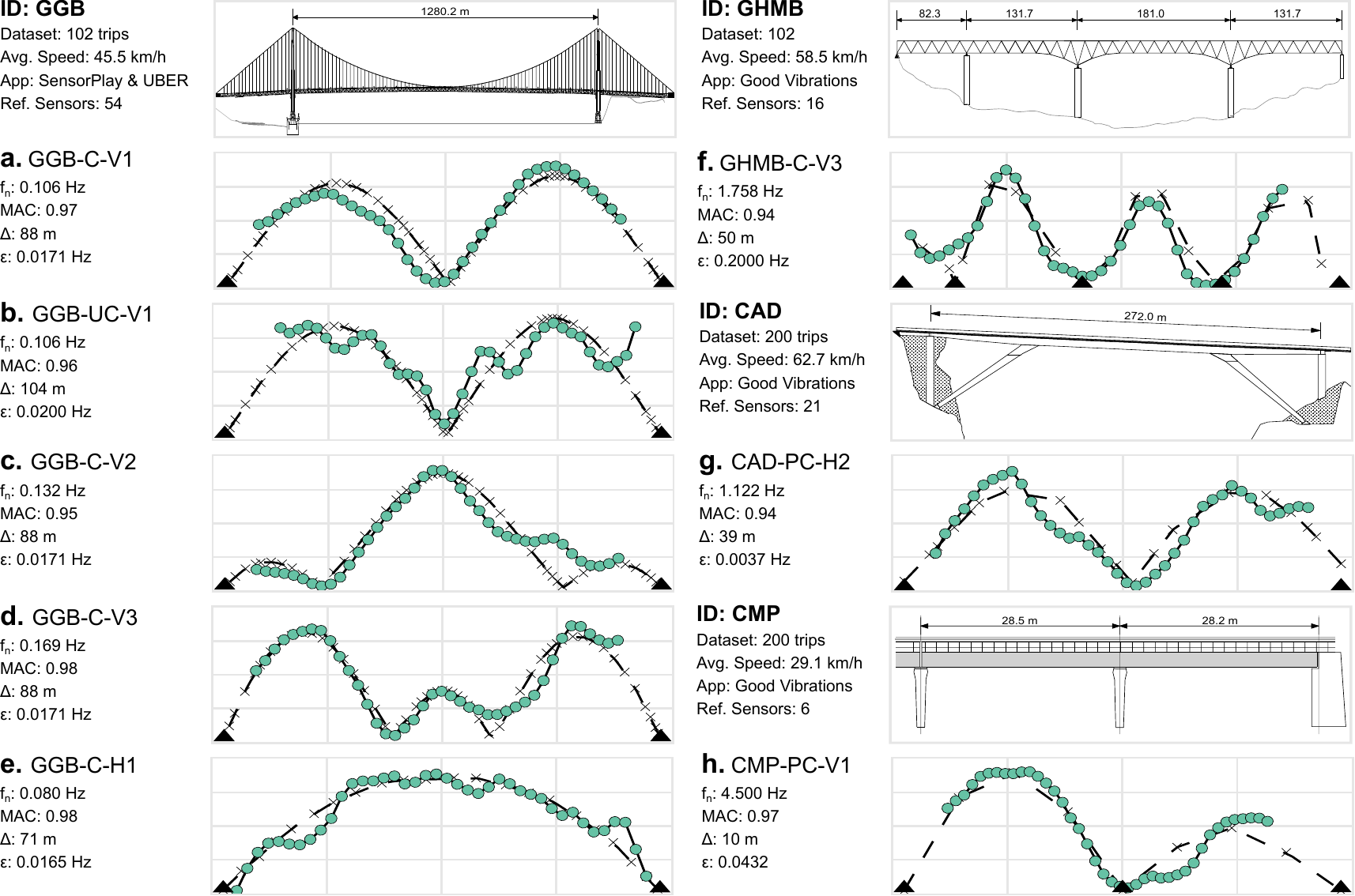}
\caption{AMS estimates for all case studies: (a-e) Golden Gate Bridge, (f) Gene Hartzell Memorial Bridge, (g) Cadore Bridge, and (h) Ciampino Bridge. Subtitles are arranged as follows: Bridge ID | Natural Frequency | MAC Value. The bridge ID is composed of three parts. First, the bridge acronym. Then, the data collection method: controlled (C), uncontrolled (UC) or partially controlled (PC). Lastly, the mode direction (vertical (V) or horizontal (H)). The estimate from the mobile sensing data is shown with green markers and a solid black line. While, the reference is displayed with a dotted line and black X markers. Each X also represents the location of the fixed sensor on the bridge during the original study. Last, black triangles show structural supports in the in the direction of the mode.} \label{fig:mode_shapes}
\end{figure*}




This study examines the efficacy of the proposed method on four real-world bridges with distinct locations, designs, and traffic volumes. The applications considered a broad set of SVT conditions; data were collected using different vehicles, e.g., sedans, minivans, etc, in three types of environments, controlled, partially-controlled, and uncontrolled, by a variety of smartphone models, and with sample sizes ranging from 50 to 200. Details of the SVT data in each application are provided in the sections that follow and are summarized in Supplementary Material.

Consistent with prior work using mobile sensor networks, the resulting AMS have a very high spatial resolution; the applications below produce mode shapes with 50 points\footnote{In traditional SHM applications with fixed sensors, the spatial resolution is limited by the number of simultaneous sensors. A mode shape with 50 points requires a network of 50 synchronized sensors.}. All applications produced highly accurate AMS as measured by MAC values of 0.94 and above. These results provide strong evidence that support the collection and analyses of crowdsourced SVT data for determining \emph{spatial vibration characteristics} of existing bridges.

\subsection*{Golden Gate Bridge}

The landmark Golden Gate Bridge (GGB) connects San Francisco to Marin County with an annual average daily traffic (AADT) of 109,000. The suspension bridge has a main span of 1,280 m over the San Francisco Bay. Due to its structural flexibility, the modal frequencies are much lower than the other bridges in this study. The first three vertical modal frequencies are 0.106, 0.132 and 0.169 Hz \cite{pakzad2009statistical}, and the fundamental transverse mode has been estimated within 0.05 and 0.095 Hz by \cite{abdel1985ambient} and \cite{ccelebi2012golden}. For SVT data, two distinct datasets are evaluated. The first dataset called GGB-C consists of 102 trips collected in a controlled environment \cite{matarazzo2020crowdsourcing}, i.e., key variables such as vehicle velocity and smartphone orientation were controlled. The second SVT dataset GGB-UC was provided by Uber from its ride-hailing fleet, which consists of 50 trips made by a diverse set of drivers and vehicles (thus, uncontrolled). 

The methodology is implemented as described in Table \ref{tbl:alg}. In summary, four modes were identified from the controlled data (GGB-C): three vertical and one transverse. In addition, the first vertical mode was identified from the uncontrolled data (GGB-UC). MAC values were calculated with respect to the locations of the fixed sensors from prior work \cite{pakzad2009statistical} (42 locations for vertical modes and 19 for the transverse mode). Impressively, the MAC values for all AMS estimates for both SVT datasets are $0.95$ or above as noted in Figure \ref{fig:mode_shapes}a-e. Furthermore the identification of a \emph{transverse} mode using SVT data is unprecedented, and the corresponding AMS was found with a MAC value of $0.98$. Lastly, it is worth noting that the AMS of the first vertical mode was identified fairly well from the uncontrolled data as indicated with a MAC value of $0.96$ and shown in Figure \ref{fig:mode_shapes}b.

\subsection*{Gene Hartzell Memorial Bridge}
The Gene Hartzell Memorial Bridge (GHMB) spans the Lehigh river along route 33 in Easton, PA, United States with an AADT of 59,000. The bridge structure is a steel truss with four spans and a total length of 540 m with a longest span of 180 m. The first three vertical modes are found to be 0.87, 1.34 and 1.78 Hz. The SVT data called GHMB is a controlled dataset of 332 trips  collected by the research team's vehicles and the \textit{Good Vibrations} App (more information in Materials and Methods section). For data collection, the team drove at a set of prescribed speeds detailed in the Supplementary Material. The mean speed across all trips was 82.25 km/h\footnote{the speed limit on the bridge is 104.6 km/h}.

For analysis, the dataset was subdivided to consider the slowest 102 trips. The idea of dividing large datasets into potentially ``more informative'' subsets is considered further in the following section. With this subset, the AMS for the third vertical was identified (Figure \ref{fig:mode_shapes}f). The corresponding MAC value was $0.94$, which was calculated using 14 reference locations. The first two vertical modes were not reliably identified, emphasizing that some vibration modes may not be observable in certain SVT subsets.

\subsection*{Cadore Bridge}
The Cadore Bridge is a 272 m long, steel-rigid-arch structure in northern Italy with an AADT of 20,000. The bridge was inspected and rehabilitated in 2011 and the fundamental natural frequencies are reported as 0.68, 1.24, and 1.80 Hz for the first three vertical modes. In addition, the bridge was inspected again with a fixed sensor network in 2021 and vertical and transverse model properties are extracted (see Supplementary Material). 

The SVT data called CAD-PC consists of 884 partially-controlled samples recorded using the \textit{Good Vibrations} app by Anas Sp.A operators. The vehicle speeds are in average $62.7$ km/h. Of the full dataset, the 200 slowest trips were selected for further analysis.  With this subset of data, Figure \ref{fig:mode_shapes}g displays the second transverse mode is found at a frequency of 1.122 Hz. The MAC value between the estimated AMS and the reference shape (17 locations) was $0.94$ .

\subsection*{Ciampino Bridge}
This bridge is one segment of an elevated intersection located in the Ciampino district of Rome, Italy. The AADT for this bridge is approximately 82,000. The segment consists of two adjacent continuous spans of reinforced concrete with the total length of 56.7 m. Since September 2020, the bridge has been monitored using a fixed sensor network with the first two natural frequencies identified as 4.5 and 6.8 Hz. The SVT data called CMP-PC consists of 992 partially-controlled datasets collected by Anas Sp.A. using the \textit{Good Vibrations App}.
The results shown in Figure \ref{fig:mode_shapes}h are from the aggregation of the 200 samples with the lowest speeds. The identified AMS in Ciampino bridge has a MAC of $0.97$ when compared to the reference at six locations.

\subsection*{Study of data quantity and quality}
\begin{figure*}[ht]
\centering
\includegraphics{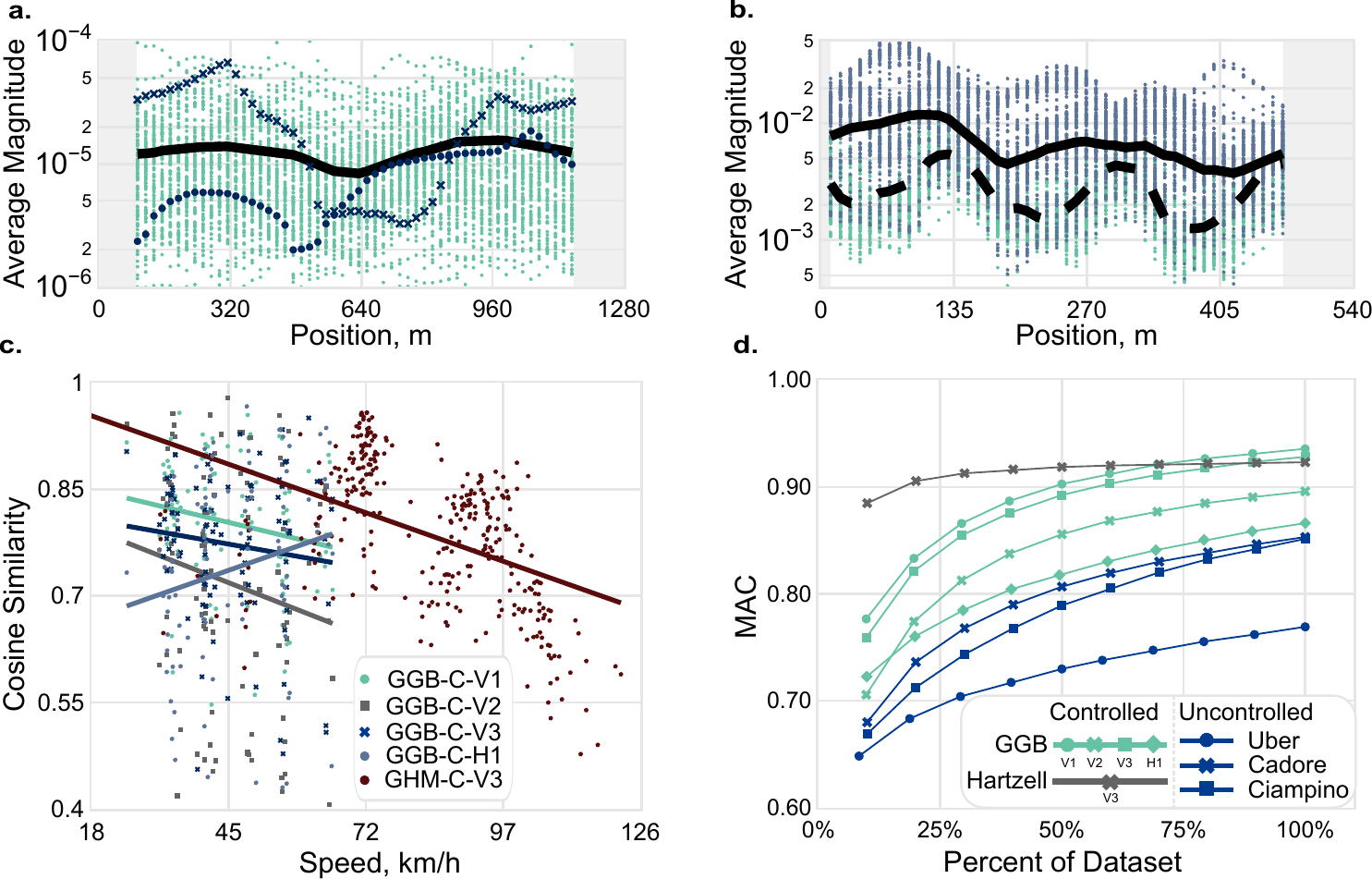}
\caption{Analysis of the dataset size and quality of samples. (a) the estimated AMS from GGB-C-V1 - shown in black - are compared with spatial patterns from single samples - shown with green markers. Two random trips are highlighted as blue markers. Marker represents the mean magnitude from a segment of the bridge, which corresponds to the absolute value of the mean of $B_{ij}$ from Table \ref{tbl:alg}. (b) The GHMB dataset is filtered by speed and the 231 fastest speeds are shown in blue and the 102 slowest speeds in green. The solid black line is the mean with respect to the full dataset, while, the dashed black line is the mean of the slowest 102 - Figure \ref{fig:mode_shapes} GHM-C-V3 result. (c) Analysis of the effect of speed on the quality of a single trip. The cosine similarity between a mode shape reference and the averaged magnitudes from a singe trip is used as a proxy metric for quality. The cosine similarity is calculated for all controlled cases and modes, and a liner trendline is fit for each mode. (d) Analysis of the effect of dataset size on the accuracy of the estimate. In this analysis, the full dataset was subsampled with replacement to create 100,000 bootstrapped datasets, and each bootstrapped dataset was processed with the averaging scheme. Each point on (d) represents the mean MAC from the 100,000 bootstrapped datasets at their respective percent of dataset size. }
\label{fig:accuracy}
\end{figure*}


The AMS results for the variety of bridges studied present an opportunity to broadly assess trends and sensitivities with respect to individual trips or groups of trips, i.e., subsets. Although many factors affect how the bridge signal is transmitted from the road surface to the cabin, this section focuses on the influence of an important feature identified in prior work: vehicle speed. 

GGB-C is a \textit{controlled} dataset containing a lower average vehicle speed compared to other datasets. Prior work discusses how controlled datasets generally exhibit higher SNR which lead to the identification of a relatively larger number of modes \cite{matarazzo2020crowdsourcing}. This is consistent with our findings; this study extracted the highest number of AMS (four) from this controlled dataset compared to other datasets. Conversely, datasets with high speeds are attributed to lower SNR as it reduces the time of data collection and intensifies the vehicle dynamical response to the road irregularities \cite{lin2005use,malekjafarian2017use}. 

For the Ciampino Bridge and the Cadore Bridge, the sample sizes of full datasets are significantly larger than sample sizes used in this study. The subsets used in the analyses are based on the premise that the sensing speed is a strong predictor for SNR, which is a proxy for signal ``quality''. Nonetheless, it is important to test this hypothesis by evaluating the  sensitivity of the AMS results to the vehicle speed and the size of dataset. These considerations become especially important when data volumes become orders of magnitude larger, i.e., tens of thousands of datasets. For instance, crowdsensing platforms with monthly data streams on the order of millions would greatly benefit from pre-processing tools that can filter out datasets that may be noisy or less likely to positively impact SHM features of interest. This process could operate by flagging datasets based on metadata analyses, e.g., speed, smartphone, vehicle type, etc., that strongly correlated with less accurate AMS.


Figures \ref{fig:accuracy}a-c break down results with respect to the vehicle's speed. Figure \ref{fig:accuracy}a is a scatter of the contributions to the AMS (mode V1 in GGB-C) from each trip across the Golden Gate Bridge, two of which are highlighted in blue as examples. The average of all samples is depicted as the AMS which is presented by dark continuous line. In a crowdsourced data collection campaign where results from large number of samples are averaged, it is possible to achieve accurate results even if a few individual trips are excessively noisy. In other words, while all datasets may be weighted equally during aggregation, they do not equally contribute to structural dynamics information. Figure \ref{fig:accuracy}b highlights the band-passed frequency-space plot (mode V3 in GHM-C) for two speed subsets, where blue points represent faster trips and green slower trips, respectively. Overall a stark difference is visible in Figure \ref{fig:accuracy}b where trips with higher vehicle speeds generally possess larger magnitudes. 








Some data subsets correspond to more accurate AMS estimates than others and may be considered to be of higher quality. Cosine similarity was used to evaluate the accuracy of an AMS from a single trip in comparison to the reference AMS. Figure \ref{fig:accuracy}c plots cosine similarity versus vehicle speed for five different modes over two bridges (GGB and GHMB). The GHMB dataset, red scatter in Figure \ref{fig:accuracy}c, spans the widest range of speeds and the most samples of the controlled datasets; the regression line was highly significant ($F = 80.15$, $p < .0001$), indicating an inverse dependency of sample quality on speed. While maintaining an inverse relationship, the fits for the GGB modes are not significant ($p > .05$). The fit for the horizontal GGB mode has a positive slope, suggesting that mode shape accuracy improves with vehicle speed, which is inconsistent with our hypothesis. Overall, these results are preliminary and provide limited quantification on how vehicle speed can affect quality of some specific mode shapes. Complications arise from large difference in the range of vehicle speeds considered in the GGB and GHMB applications, 20-70 km/hr vs. 20-120 km/hr, as well as the nonuniform sampling of the speeds. In summary, while we can confirm speed is a quality predictor in one case, i.e., GHMB, this observation is not statistically significant for GGB modes. The quality of an individual SVT is affected by a wide range of features including bridge length, traffic speed, length ratio, road profile roughness, expansion joints, to name a few. In our analysis, we found evidence of speed affecting a metric of mode shape quality, yet further investigation is needed to establish empirical rules for predicting the quality of individual SVT data based on trip metadata and features.



In Figure \ref{fig:accuracy}d, the relationship between sample size and MAC value is presented for the identified AMSs. The sample size is presented in terms of percentage of the entire sample set. For each percentage level, 100,000 minibatches were randomly selected with replacement from the full dataset (i.e., sample bootstrapping) and the average MAC value was calculated. From Figure \ref{fig:accuracy}d, in all cases, without exception, the accuracy of identified AMSs monotonically increased with sample size. The extent of this improvement varied with each mode and with respect to the controllability of the SVT data. In one case, increasing sample sizes rose the MAC value of the first mode from Uber data ($N=50$) by 31\%. There are two further observations: (1) the rate at which accuracy increases decreases with sample size, and (2) accuracy increases at a slower rate for the uncontrolled data. This suggests that when compared to uncontrolled trips, fewer controlled trips would be needed to achieve a certain AMS accuracy level for a given mode. For instance, consider the first vertical mode of the GGB: the MAC value from $50 \%$ of the controlled  data (50 samples used) is about $0.90$, which is considerably higher than $0.75$, the value from the Uber data (50 samples used).

\section*{Discussion}

Four studies in this paper demonstrated a wide range of real-world examples in which important spatial vibration characteristics of bridges, i.e., absolute value mode shapes (AMSs), can be extracted successfully from crowdsourced smartphone-vehicle trip (SVT) data. Bridges from three general classes were considered: short-span, medium-to-long-span, and long-span, and three types of crowdsourced SVT datasets were used:  controlled, partially-controlled, and uncontrolled. 

The successful applications on the Golden Gate Bridge (GGB) show capabilities for identifying AMSs of long-span bridges in both vertical and horizontal directions (see Figure \ref{fig:mode_shapes}a-e). Bridges with long main spans ($>500$ m) belong to a small, yet special category. While there are fewer than 100 such bridges in the world, they have a high structural flexibility which is tied to low modal frequencies (i.e., several modes below $1$ Hz \cite{pakzad2009statistical,chen2013handbook}). This is notable for the case of SVT data as this frequency range is distinctly below the typical frequency range of vehicle suspension systems ($1-3$ Hz). Thus, negative effects of vehicle-bridge-road interaction are mitigated. Additionally, for a given speed and sampling rate, SVT data sets collected on long-span bridges will possess a larger sample size. Finally, long-span bridges are among the most travelled, as measured by large AADT values, which facilitates high-rate SVT data streams. In summary, with the potential for large volumes of high quality datasets, SVT data may present their highest value to long-span bridges.



The studies on the Cadore Bridge (CAD) and Gene Hartzell Memorial Bridge (GHMB) confirmed applicability of the method to a larger group of existing bridges with maximum spans between $50-500$ m, i.e., medium- to long-span bridges. According to the national bridge inventory (NBI) curated by the federal highways administration (FHWA), about $12,000$ bridges ($2 \%$) in the US belong to this class. Similarly, the successful study on the short-span Ciampino Bridge (CMP) represents applicability of the method to US bridges having maximum spans between 15 m and 50 m, which includes nearly $163,000$ bridges ($26 \%$).

From a structural dynamics perspective, bridges with shorter spans are stiffer than long-span bridges and are more likely to possess some modal frequencies between the typical vehicle band $1-3$ Hz. Note that the modal frequencies for both the second horizontal mode for CAD (see CAD-PC-H2 in Figure \ref{fig:mode_shapes}g) and the third vertical mode for GHMB (see GHMB-C-V3 in Figure \ref{fig:mode_shapes}f) fall within this range and are therefore subject to high noise from vehicle-bridge-road interaction. This methodology does not include modeling of the vehicle system nor any processes for decoupling vehicle dynamics, e.g., empirical mode decomposition, deconvolution, etc. \cite{eshkevari2020bridge}; nonetheless, accurate AMSs were extracted, which may suggest that large and diverse SVT data sets, acquired using many different vehicle, help mitigate negative interaction effects. It is notable that only one AMS was successfully extracted from each of these bridges. This result is partially dictated by the random nature of the traffic excitation: there is no guarantee that a given mode is sufficiently active in the measured structural response at the space-time coordinates of the SVT data. Large data volumes (on the order of thousands or hundreds of thousands) would dramatically improve the odds that individual structural modes have a strong presence in the aggregate space-frequency maps. Generally, for robust damage detection, several bridge modes should be consistently tracked over time.

The AMSs are directly linked to reliable indicators of structural damage, such as mode shape curvature, total modal assurance criteria (TMAC), etc. \cite{gao2002damage}, and thereby establish a fundamental functionality for a damage detection system. Robust synthetic analyses in \emph{Supplementary Note 1} exemplify two crowdsensing scenarios in which these features are used to detect the presence of bridge damage. A 60-m long bridge was considered with a mobile sensor network that produced 100 SVT datasets per day over the course of six months. The identified AMSs for three modes were used to track the TMAC value over time (days) which led to an accurate damage detection within a few days of the damage event.


The approach proposed in this paper is based on statistical signal processing techniques and does not utilize machine learning (ML). Over the past decade, applications of data-driven ML methods have initiated enormous advances in the sciences, broadly shaping fields such as physics, computer vision, robotics, natural language processing, neuroscience, etc. \cite{jordan2015machine,marx2013big,he2016deep,brunton2016discovering}, with growing applications in civil engineering \cite{lee2005neural,spencer2019advances, khaloo2018utilizing, rafiei2018novel}. That is, the results here establish the initial, known capabilities for extracting spatial vibration characteristics features but may underreport the extent of dynamic features and accuracies that are possible for the considered data sets. For instance, in the studies presented, the accuracy of the results from SVT data were evaluated through a direct comparison with results based on traditional fixed sensor networks, while trained AI models may enable techniques for automated cross-validation of modal properties.

Bridge condition monitoring based on SVT data leverages existing vehicle mobility networks and high smartphone penetration rates. These approaches offer very low upfront monitoring costs compared to traditional sensing systems and enable a scalable information retrieval process: all bridges that are covered by the smartphone-vehicle network can be analyzed using these tools. Urban bridges generally have high AADT levels, such that low participation rates can still yield ample data streaming volumes, e.g., $1\%$ of the AADT of the Golden Gate Bridge amounts to over $1,000$ daily trips. Conversely, data streaming volumes for bridges in rural areas may be limited by comparatively lower AADT values; yet, large-scale and long-term data sets combined with advances in transfer learning could help utilize prior analyses of similar bridges to enable dynamic property estimation using sparse SVT data.

A key ambition for crowdsourcing infrastructure condition information is the creation of an open database of bridge vibration records, curated with historic dynamic property estimates, bridge model files, design details such as material or span, etc. The overarching goal is to provide bridge owners and inspectors an opportunity to focus on key, pre-identified bridge locations or components of interest, to enable more frequent, more quantitative inspections, to support preventative maintenance protocols. Such a database would complement existing archives, e.g., the National Bridge Inventory, which is managed by the FHWA, with the addition of maintaining historic data and records. Regular observations of structural behavior over a long term (over one year) are essential to establishing baselines and performance benchmarks needed for condition evaluations \cite{smith2016studies}. Centralization and standardization of archival vibration data and structural health reports would be critical to the success and impact of this program. First, a widespread coverage of bridges would promote participation of bridge owners. Second, standardized and curated data sets assist with the rapid development and application of new techniques, such as developing trained AI models for condition monitoring. 

The establishment of a mobile sensing platform of this scale creates new interdisciplinary challenges. The everyday use and reliability has substantial computing, network, and cloud infrastructure needs for which existing cellular networks and smartphone ownership levels provide an important foundation. There are a very large number of eligible SVT datasets that extensively cover US infrastructure and have already been collected by ridesharing companies and apps such as Uber, Lyft, Waze, etc. Fig. \ref{fig:mode_shapes}b shows a successful implementation based on a small number of uncontrolled SVT trips collected by Uber. There is high potential value within these enormous, pre-existing datasets, which present opportunities for ridesharing businesses to collaborate with municipalities and state Departments of Transportation. Advances in smart vehicle technologies, e.g., vehicle-to-infrastructure communication, could made a large positive impact by accelerating data transmission, storage, and analysis. Developments in material science can help connect sparse structural response measurements to local and global behavior \cite{feng2022visual}, e.g., load distributions \cite{guo2020semi}, multi-scale analysis and design \cite{nadkarni2014dynamics}. The concept of an open database of bridge records poses potential legal constraints and questions which will depend on jurisdiction. Is there a need to restrict database access to a limited number of users, organizations, or institutions for national security? Would there be a there a legal obligation for a bridge owner to act on data-driven reports describing significant indicators of severe structural damage? 

The findings of this paper could have a substantial impact on the monitoring and management of bridges globally. The tools developed are broadly applicable to all bridges; yet, there remains a need to continue to study the accuracies and efficacies of using SVT data in additional scenarios to better identify strengths and potential limitations of the approach for existing bridges.  The results presented in \emph{Supplementary Note 1} emphasize the power of regularly collecting and analyzing SVT data over time for damage detection and support the integration of low-cost, large-scale monitoring into bridge management practices.

\subsection*{Data Collection and Processing}


The \textit{Good Vibrations} Android application has been developed by the researchers at the MIT Senseable City Lab to record vibrations, locations, and orientations data generated when a smartphone is moving over a monitored infrastructure. The main objectives of this application can be summarized as follows:

\begin{enumerate}
    \item Timely detect when a smartphone is entering or leaving a monitored area.
    \item Record and store the data coming from the sensors of the smartphone at the highest available frequency.
    \item Minimize the energy consumption of the application.
\end{enumerate}

However, the first and last item of this list are generally in contrast. On one hand, to detect the position of a smartphone with a good level of accuracy while tracking its movements it is necessary to keep the GPS receiver always ON; on the other hand an active GPS receiver can consume most of the battery of a smartphone. To solve this issue we decided to implement a multi-barrier activation approach. 
The first activation barrier consists of detecting if the user is currently driving. When the application is active, a service listens for an activity change event generated by the Android OS. Activity detection is a service provided by the Android OS which allows, thanks to a Machine Learning model, to detect if the smartphone is standing still or carried by a user who is walking, biking, or driving by just analyzing the signal generated by the IMU sensors of the smartphones. This approach is more energy conservative compared to other localization methods at the cost of a small activation delay \footnote{ https://developers.google.com/location-context/activity-recognition/}. The second activation barrier can be activated only if the user is driving close to a monitored infrastructure. In this case the application relies on the GPS receiver of the smartphone and it implements the following logic to detect if the user is \emph{close}: If a monitored infrastructure is reachable in less than 60 seconds the GPS receiver is left ON and the current location of the device is collected every 5 seconds. Otherwise, the application will turn off the GPS module and will schedule a new activation of the GPS receiver in $t$ seconds. This $t$ value is computed based on the distance from the closest infrastructure.

The $t$ value is a very conservative estimate, but in our experience it is the best trade-off between accuracy, energy consumption, and computation capabilities.
Finally, if the user is driving, the location and speed reported by the GPS receiver indicate that a monitored infrastructure is reachable in less than 30 seconds, and the accuracy provided by the GPS receiver is less than 10 m the recording process can start.
The data generated by the GPS receiver, (timestamp, latitude, longitude, speed, accuracy) the rotation vector (timestamp and rotation quaternion) and the accelerometer (timestamp and x, y, z components of the acceleration vector) are recorded independently at the maximum sample rate allowed by the device. Generally, GPS data is recorded every second, while Accelerations and Rotation data rate can vary depending on the specific device between 50 and 500 Hz. 

To allow for the fastest possible recording, these data are initially stored in memory and then moved to the storage of the device when the data collection is over (or periodically to avoid filling up the memory). To avoid losing measurements during the memory off load, the entire collection process is multithreaded and based on synchronization queues in a producer-consumer fashion. The new data is inserted into dedicated queues (one for each sensor) and a writer thread is responsible for moving the data from the queues to the physical storage of the device when required. This minimizes the amount of code executed as a response to a new data sample, while leaving the data in terms of timestamps and values exactly as provided by the OS. Once the user leaves the monitored area, a first check on the GPS trace is performed. However, a driver could get really close to an infrastructure without crossing it. To filter out these cases the application checks that the recorded GPS trajectory intersects a set of checkpoints located on the bridge. If this happens, the scan is ready for uploading, otherwise the scan is discarded. Before uploading the collected scans to the Cloud, the data is compressed and divided into smaller chunks to ease the upload process. Finally, when the smartphone is connected to the Internet the scans are uploaded to the Cloud and if the process is successful the uploaded data is removed from the smartphone. 

\subsection*{Absolute Modeshape Identification Methodology}

\begin{table*}
\centering
\fbox{\begin{minipage}{0.75\paperwidth}
\begin{enumerate}
    \item Depending on the axis of interest, corresponding raw acceleration signals $a^{r}_i(t)$ are collected with smartphones: $i=1,2,...,N$. 
    \item Preprocessing: $a^{r}_i(t)$ is downsampled to a desired frequency range, yielding $a_i(t)$. The downsampled signals are checked for stationarity using augmented Dickey-Fuller test \cite{dickey1979distribution} to filter out corrupted signals with baseline shifts or unreasonable variations.
    \item (Optional) Depending on the direction of the vehicle motion (right-left or left-right) and in case that the inspected span is surrounded by expansion joints, a fractual value $\alpha \in [0,0.2]$ is selected and the initial portions of the signals are trimmed: $a_i(t):=a_i(t_n): n >\alpha \times |a_i(t)|$.
    \item Synchrosqueezed wavelet transform is calculated using $a_i(t)$, resulting $T_{a_i}(f,t)$, in which $0\leq f \leq f_{Nyq}$ and $f_{Nyq}$ is the Nyquist frequency defined by the sampling rate.
    \item $T_{a_i}(f,t)$ is mapped into spatial coordinates using available time-stamped GPS coordinates (provided by the app), resulting $T_{a_i}(f,x)$ in which $x \in [0,L_{br}]$ is the longitudinal position on the bridge with total length of $L_{br}$.
    \item Select values for bandwidth $\epsilon$, spatial segment width $\Delta$, and a spatial stride $S$. 
    \item Given a desired natural frequency $f_k$, a series of frequency-space grids are defined as follows:\\
    \rule{0pt}{4ex}
    $B_{ij}=T_{a_i}([f_k-\epsilon/2,f_k+\epsilon/2],[j\times S, \max\{L_{br}, \Delta + j\times S\}])$ for $j=1,...,\lfloor(L_{br}-\Delta)/S\rfloor+1$. \\
    \rule{0pt}{4ex}
    $B_{ij}$ contains wavelet amplitudes for signal $a_i(t)$ in a $j^{th}$ frequency-space grid. If $|B_{ij}| = 0$, it is skipped.
    \item The mean of amplitudes over all signals for each frequency-space grid is calculated: $\mu_j = \frac{1}{N}\sum_1^N{p}: p \in B_{ij}$. 
    \item Finally, an ordered set of $\mu_j$ for $j=1,...,\lfloor(L_{br}-\Delta)/S\rfloor+1$ presents the aggregated absolute spatial pattern of the bridge on frequency $f_k$. For a fair choice of $f_k$, this spatial signature should converge to the absolute natural mode shape.
    
\end{enumerate}
\end{minipage}}
\caption{Summary of the core methodology used for identifying absolute natural mode shapes.}\label{tbl:alg}
\end{table*}

The algorithm detects the change in magnitude of known modal frequencies with the position on the bridge, which is proportional to the AMS. For data collection, vehicles carry smartphones that take acceleration and location measurements. Since the signals are collected within moving vehicles, the bridge vibrations are highly contaminated by the road profile and vehicle dynamics. Furthermore, the low quality smartphone sensors introduce additional sampling errors and measurement noise. For these reasons, the bridge vibrations are imperceptible in a single signal, however, by averaging many trips the spatial characteristics of the vibrations appear.\par

Due to the noisy nature of the data collection process, the datasets are filtered proior to aggregation. Under visual inspection of the collected signals in the uncontrolled datasets, noticeable trends are apparent, and as shown in the analysis in Figure \ref{fig:accuracy}, the speed of the vehicle has a detrimental effect on accuracy. For these reasons, the datasets are filtered first based on speed, and second, with the augmented dickey fuller (ADF) test for signal stationarity. \par

The aggregation process spatially averages frequency contents to determine the relative magnitude differences at each location on the bridge. To achieve this, the bridge is divided into overlapping segments; the width, $\Delta$, and stride, $S$, (distance segment centers) are two parameters of the method. The synchrosqueezed wavelet transform converts each acceleration time series to the time-frequency domain, and the instantaneous magnitudes are calculated. Lastly, averaging the magnitudes in each segment, $B_{ij}$, at the modal frequencies of the bridge for all trips yields the AMS. Including a small bandwidth, $\epsilon$, around the modal frequency in the averaging process leads to robust results on noisy datasets.\par

The method relies on the prior knowledge of the modal frequencies, and for this study, the natural frequencies are determined in two different ways. First, the frequencies are found using a traditional fixed sensor network, the state of practice in SHM, which were used as a reference for analysis. Second, the natural frequencies are found with the recently proposed algorithm Most Probable Modal Frequencies (MPMF) \cite{matarazzo2020crowdsourcing}. The goal of determining the frequencies in such a way is to demonstrates that all of the results in the work can be found only using measurements obtained from passing vehicles. The results from this analysis are found in the Supplementary Material.\par

\bibliographystyle{unsrt}  
\bibliography{references}

\end{document}